# Global Distribution of the Key Species on the Surface of Europa

Jiazheng Li[1,2], Yinsi Shou[2], Cheng Li[2], and Xianzhe Jia[2]
[1] Space Science Institute, Macau University of Science and Technology, Avenida Wai Long, Taipa, Macau, People's Republic of China; jzli@must.edu.mo
[2] Department of Climate and Space Sciences and Engineering, University of Michigan, Ann Arbor, MI, USA

## Abstract

The icy surface of Europa is continuously bombarded by ions and electrons from Jupiter's magnetosphere. The bombardment of the particles dissociates water molecules on the surface of Europa and introduces impurities to the icy surface. Such processes lead to the generation of the nonwater species on the surface of Europa. These chemical species are closely related to the chemistry of the icy crust and the subsurface ocean, as well as Europa's habitability. However, our knowledge of the global distribution of these species is limited due to the sparse satellite and telescope observations on Europa. In this study, we combine a Europa plasma model and a chemical-transport model to simulate the global distribution of the key nonwater species on the surface of Europa. The initial results from our model agree well with the existing observations on the distributions of $H_2SO_4$ and $SO_2$, but they show a significant discrepancy with the observed distribution of $H_2O_2$. Sensitivity tests on the reaction rate coefficients indicate that the simulated global distribution of all three species fits the observations well if the reaction rate coefficients in the ice are reduced by 1 order of magnitude. This finding provides a useful constraint on the rate coefficient of the chemical reactions in the ice. Furthermore, our model predicts that the $O_2$ on the surface ice of Europa is concentrated on the leading hemisphere. The simulated global distribution of the key species on Europa may provide useful guidance for future missions to Europa, such as Europa Clipper and JUICE.

*Unified Astronomy Thesaurus concepts:* Europa (2189); Galilean satellites (627); Surface composition (2115); Planetary surfaces (2113); Surface processes (2116); Surface ices (2117); Astrobiology (74); Planetary magnetospheres (997); Jovian satellites (872)

## 1. Introduction

The continuous bombardment of Europa's surface by charged particles (including electrons and hydrogen, oxygen and sulfur ions; J. F. Cooper et al. 2001) from Jupiter's magnetosphere is one of the major driving forces of the surface processes on Europa. Energetic ions can release $H_2O$ and $H_2O$ products ($H_2$, $O_2$, etc.) from the surface ice via sputtering and radiolysis processes to space, which lead to the formation of Europa's tenuous atmosphere (W. L. Brown et al. 1982, 1984; A. Bar-Nun et al. 1985; W. H. Ip 1996; W. H. Ip et al. 1998; W. H. Smyth & M. L. Marconi 2006; R. E. Johnson et al. 2009; L. Roth et al. 2016; B. D. Teolis et al. 2017). Ion and electron irradiation on the surface ice can dissociate water molecules and produce radicals, which contribute to the chemical alteration of the ice and the production of oxidants (R. W. Carlson et al. 1999a; R. A. Baragiola et al. 2002; J. R. Spencer & W. M. Calvin 2002; T. M. Orlando & M. T. Sieger 2003; K. P. Hand et al. 2006; W. Zheng et al. 2006; K. P. Hand & M. E. Brown 2013; A. Galli et al. 2018; T. A. Nordheim et al. 2018; A. Vorburger & P. Wurz 2018; R. M. Meier & M. J. Loeffler 2020; M. R. Davis et al. 2021; J. Li et al. 2022). Originated from the volcanoes on Io, the sulfur ions in Jupiter's magnetosphere bring sulfur to the surface ice of Europa and facilitate the production of $SO_2$, $H_2SO_4$, and other sulfur-containing species in the ice (R. W. Carlson et al. 1999b, 2002; G. Strazzulla et al. 2007, 2009; M. J. Loeffler et al. 2011; J. J. Ding et al. 2013;

M. S. Gudipati et al. 2021; T. M. Becker et al. 2022; J. Li & C. Li 2023).

Studies have shown that due to the influence of the magnetic field around Europa on the trajectories of the magnetospheric charged particles, the incoming fluxes of the charged particles on Europa's surface are not uniformly distributed (T. A. Nordheim et al. 2018, 2022; Y. Shou et al. 2025). Therefore, the concentrations of the nonwater species generated by the bombardment of Europa's surface are different from place to place. A few studies have used satellite and telescope observations to determine the global distribution of those species. Galileo Near-Infrared Mapping Spectrometer observations show that sulfuric acid and its hydrates are abundant on the surface of Europa and likely to be found in Europa's trailing hemisphere rather than its leading hemisphere (R. W. Carlson et al. 2005; J. B. Dalton et al. 2013). T. M. Becker et al. (2022) use the midultraviolet channel of the Hubble Space Telescope Imaging Spectrograph to map the global surface distribution of $SO_2$ on Europa. The 280 nm feature, which is attributed to $SO_2$, is concentrated toward the apex of the trailing hemisphere of Europa. As for the oxidants, R. W. Carlson et al. (1999a) show that the surface hydrogen peroxide concentration is about 0.13% in the leading anti-Jovian quadrant based on Galileo Near-Infrared Mapping Spectrometer observations. A follow-up study, K. P. Hand & M. E. Brown (2013), uses KECK II observations of Europa to determine the $H_2O_2$ abundance on the surface of Europa. The observations show that only a small amount of $H_2O_2$ exists on the anti-Jovian and sub-Jovian hemispheres of Europa, while almost no hydrogen peroxide is detected during observations of just the trailing hemisphere. S. K. Trumbo et al. (2019) find that the largest hydrogen peroxide absorptions are located at low latitudes on the leading and anti-Jovian hemispheres of Europa based on the KECK II observations in 2016 and 2018. A recent study, P. Wu et al. (2024), obtains the similar distribution with the latest KECK observations. Although these studies provide us with the basic information about the distribution of the key nonwater species on Europa, they are limited by the sparse and low spatial resolution observations that only cover a part of Europa's surface. Meanwhile, there are no observations on one of the most important species on Europa, $O_2$. Since the key chemical species on the surface of Europa are closely related to the chemistry of Europa's icy crust and the compositions of its subsurface ocean, comprehensive studies on the global distribution of the nonwater species on Europa are highly necessary. Such work may also provide useful guidance for future missions to Europa.

In this paper, we make the first attempt to simulate the global distribution of the important nonwater species on the surface of Europa. Two models, a Europa plasma model and a chemical-transport model, are combined to achieve such a goal. The Europa plasma model is used to determine the incoming fluxes of the charged particles at each location on Europa, and the chemical-transport model makes use of the incoming fluxes to calculate the abundances of the nonwater species at each location on the surface of Europa. The Europa plasma model and the chemical-transport model are described in Sections 2 and 3, respectively. The simulated global distribution of the key nonwater species on the surface of Europa is illustrated in Section 4. In Section 5, we will draw conclusions and discuss the prospect of future studies.

## 2. The Europa Plasma Model

In this work, the plasma model is used to compute flux and energy flux distributions of various charged particles on the surface of Europa while taking the interaction between the Jovian magnetosphere and Europa's atmosphere into account, as Europa is traversing in different regions of the Jovian magnetosphere. The used plasma model combines a well-established multifluid Europa MHD (C. D. Harris et al. 2021, 2022) and a test particle model so that the test particle model is able to track the motion of charged particles of various species and at various energies in the vicinity of Europa, as the modeled particles are subject to Lorentz force by the electromagnetic (EM) fields, which can be obtained from the MHD model. We used the same plasma model and part of the results from Y. Shou et al. (2025), in which more detailed description of the plasma model is provided. Seven components of the incoming charged particles are considered in this study: thermal and energetic hydrogen ions ($H^+$), thermal and energetic oxygen ions ($O^+$), thermal and energetic sulfur ions ($S^{++}$), and electrons ($e^-$). For each component, we obtain the global distribution of its energy flux by averaging the energy flux at each location on Europa from the three simulation cases corresponding to Europa's locations in the Jovian magnetosphere at three magnetic latitudes, 0°, 5° and 10°. The incoming energy fluxes of the six major components (thermal and energetic oxygen ions, thermal and energetic sulfur ions, energetic hydrogen ions and electrons) as functions of latitude and longitude are shown in Figure 1. Although the energy fluxes of the energetic oxygen and sulfur ions (Figures 1(b) and (d)) are about 2 orders of magnitude higher than those of the thermal oxygen and sulfur ions (Figures 1(a) and (c)), their average energies are roughly 3 orders of magnitude larger. As a result, the particle fluxes of both oxygen and sulfur ions are dominated by the thermal ions.

## 3. The Chemical-transport Model

Our chemical-transport model is developed based on the Caltech/JPL chemical-transport model KINETICS, which finds the steady-state solution for a certain chemical-transport scenario (M. Allen et al. 1981; Y. L. Yung & W. B. DeMore 1999). More details of the mechanism of KINETICS can be found in G. R. Gladstone et al. (1996). Since the charged particle irradiation only affects the meter-scale ice shell near the surface, chemical transport in the horizontal direction is neglected in this study, and we use the one-dimensional chemical-transport model to simulate the chemical processes at each location of Europa. The one-dimensional KINETICS model used in this study is adopted from the special version of KINETICS presented in J. Li & C. Li (2023), which is specially designed to solve the chemical-transport problems in the scenario of icy satellites.

The basic setup of the chemical-transport model is as follows: The simulated ice shell is from the surface to the depth of 50 cm. The chemical species in our model include $H_2O$, $H$, $O$, $O_2$, $OH$, $H_2$, $O_3$, $HO_2$, $H_2O_2$, $S$, $S_2$, $S_3$, $S_4$, $SO$, $SO_2$, $SO_3$, $S_2O$, $HSO_3$, $HOSO_3$, and $H_2SO_4$. The temperature of the ice shell is fixed at 100 K. The model is driven by the irradiation-induced water dissociation and the implantation of the ions. Fifty-three possible chemical reactions are considered in the model (the reactions and their rate coefficients are listed in Table 1, which are the same as the reactions used in J. Li & C. Li 2023). Fixed-density upper boundary (surface) condition is applied to the gas-phase species listed in Table 2, while a zero-flux upper boundary condition is applied to other species. The densities of all nonwater species are fixed at 0 at the lower boundary (50 cm). The temperature of the layer is fixed at the average surface temperature of Europa (100 K).

The Jovian magnetospheric model described in Section 2 produces the incoming energy fluxes of the seven components of the incoming charged particles (thermal and energetic hydrogen ions, thermal and energetic oxygen ions, thermal and energetic sulfur ions, and electrons) on the surface of Europa as functions of latitude and longitude. For the sake of simplification, we assume the energy fluxes of the seven components result from monoenergetic particle beams. The particle energies of the monoenergetic beams are estimated from the magnetospheric model by averaging the particle energies of each component. The penetration depths of the monoenergetic beams are estimated based on the following databases: The Stopping and Range of Ions in Matter (J. F. Ziegler & J. P. Biersack 1985), PSTAR (International Commission on Radiation Units and Measurements 1993), and ESTAR (International Commission on Radiation Units and Measurements 1984). The particle energies of the monoenergetic beams and their penetration depth are listed in Table 3. During the penetration of the charged particles, their energy will be deposited in the ice and be used to dissociate water molecules. The average energy used per dissociation of one water molecule through water ionization is set at 12.7 eV. Since the dissociated water molecules may combine and reform water due to the cage effect in solids (C. G. Elles et al. 2007), we reduce the water dissociation rate by 50%, following the same approach as in J. Li & C. Li (2023). Each water molecule can be dissociated either into a H atom and an OH or a $H_2$ and an O atom. According to B. D. Teolis et al. (2009), the dissociative partitioning of the water molecule is set to be 78% for H + OH and 22% for $H_2$ + O. Meanwhile, the ions will be implanted in the surface ice. In our model, we assume that once the ions go into the surface ice, they will quickly lose their charges and be neutralized to atoms. For each monoenergetic beam, we assume that the energy deposition of the incoming charged particles and the implantation of the ions are uniformly distributed along the penetration depth.

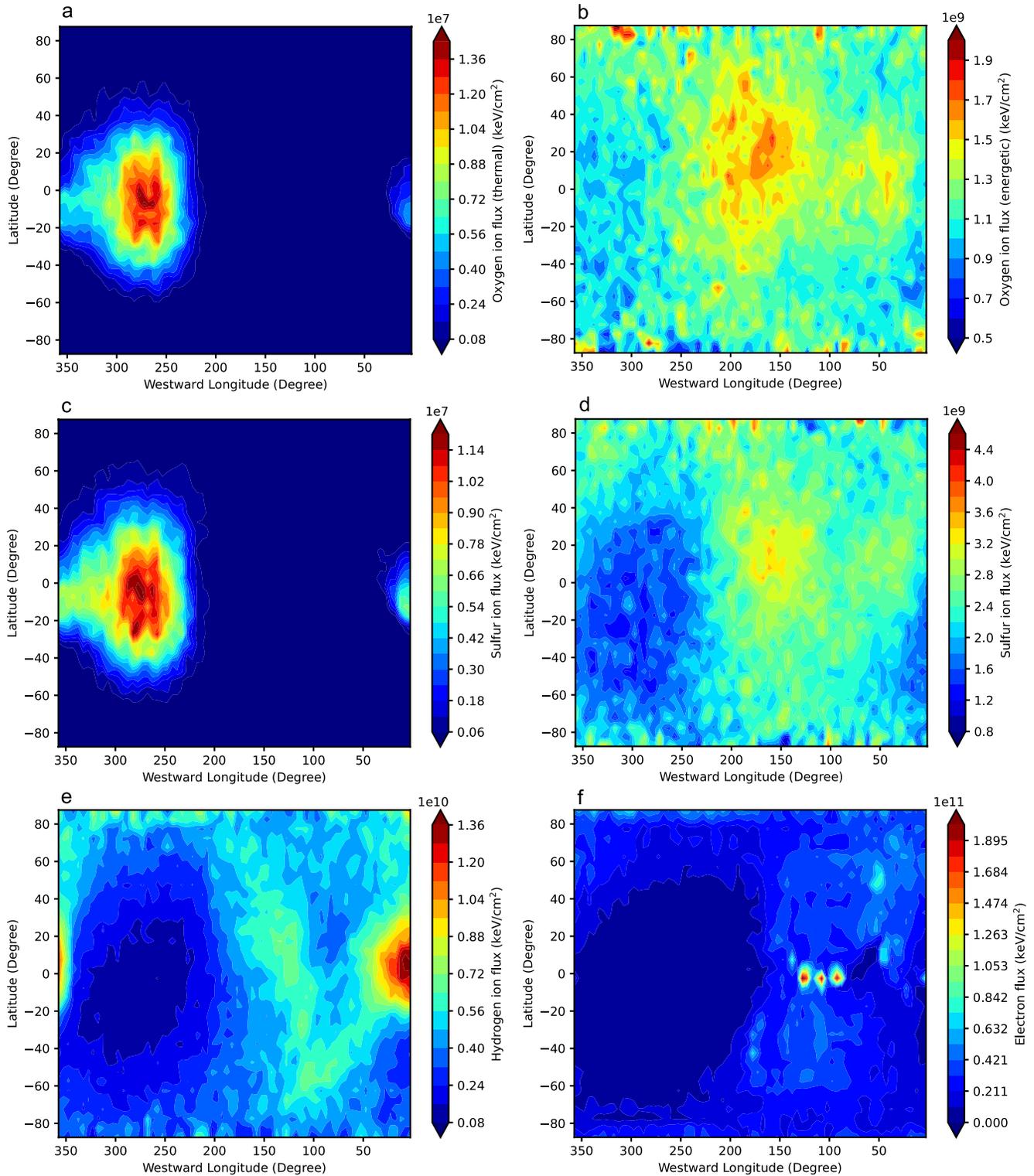

**Figure 1.** The incoming energy fluxes of the six major components on the surface of Europa as functions of latitude and longitude. (a) Thermal $O^+$. (b) Energetic $O^+$. (c) Thermal $S^{++}$. (d) Energetic $S^{++}$. (e) Energetic $H^+$. (f) Electrons.

Based on the energy fluxes of the incoming charged particles at a certain location on Europa, we are able to calculate the vertical profiles of the water dissociation rate and the ion implantation rates at this location according to the

**Table 1**
A List of Reactions Included in Our Chemical-transport Model

| Reactions | Rate Coefficients[a] |
|---|---|
| $OH + OH + M^b \rightarrow H_2O_2 + M$ | $4.21 \times 10^{-29}$ |
| $OH + OH \rightarrow O + H_2O$ | $4.63 \times 10^{-11}$ |
| $OH + H_2O_2 \rightarrow HO_2 + H_2O$ | $5.85 \times 10^{-13}$ |
| $OH + H + M \rightarrow H_2O + M$ | $3.89 \times 10^{-29}$ |
| $O + H + M \rightarrow OH + M$ | $1.30 \times 10^{-31}$ |
| $O + O + M \rightarrow O_2 + M$ | $1.21 \times 10^{-31}$ |
| $O + HO_2 \rightarrow OH + O_2$ | $2.17 \times 10^{-10}$ |
| $O + OH \rightarrow O_2 + H$ | $7.30 \times 10^{-11}$ |
| $OH + HO_2 \rightarrow O_2 + H_2O$ | $5.86 \times 10^{-10}$ |
| $O + H_2 \rightarrow OH + H$ | $4.03 \times 10^{-28}$ |
| $OH + H_2 \rightarrow H + H_2O$ | $5.84 \times 10^{-21}$ |
| $H + HO_2 \rightarrow OH + OH$ | $7.30 \times 10^{-11}$ |
| $H + HO_2 \rightarrow O_2 + H_2$ | $6.40 \times 10^{-12}$ |
| $H + H + M \rightarrow H_2 + M$ | $1.80 \times 10^{-32}$ |
| $H + H_2O_2 \rightarrow OH + H_2O$ | $1.49 \times 10^{-13}$ |
| $O_2 + H + M \rightarrow HO_2 + M$ | $1.16 \times 10^{-33}$ |
| $HO_2 + HO_2 \rightarrow H_2O_2 + O_2$ | $1.16 \times 10^{-15}$ |
| $O + O_2 + M \rightarrow O_3 + M$ | $9.33 \times 10^{-35}$ |
| $H + O_3 \rightarrow O_2 + OH$ | $2.79 \times 10^{-11}$ |
| $O + O_3 \rightarrow O_2 + O_2$ | $9.05 \times 10^{-21}$ |
| $S + O_2 \rightarrow SO + O$ | $8.06 \times 10^{-11}$ |
| $S + OH \rightarrow SO + H$ | $6.59 \times 10^{-11}$ |
| $S + O_3 \rightarrow SO + O_2$ | $1.20 \times 10^{-11}$ |
| $S + HO_2 \rightarrow SO + OH$ | $2.22 \times 10^{-10}$ |
| $S + S \rightarrow S_2$ | $1.00 \times 10^{-10}$ |
| $S + S_2 \rightarrow S_3$ | $3.00 \times 10^{-11}$ |
| $S + S_3 \rightarrow S_2 + S_2$ | $8.00 \times 10^{-11}$ |
| $S + S_3 \rightarrow S_4$ | $3.00 \times 10^{-11}$ |
| $S + S_4 \rightarrow S_2 + S_3$ | $8.00 \times 10^{-11}$ |
| $S + SO \rightarrow S_2O$ | $1.00 \times 10^{-11}$ |
| $S + S_2O \rightarrow S_2 + SO$ | $6.14 \times 10^{-18}$ |
| $O + S_2 \rightarrow SO + S$ | $9.46 \times 10^{-12}$ |
| $S_2 + S_2 \rightarrow S_4$ | $1.00 \times 10^{-10}$ |
| $S_2 + SO_3 \rightarrow S_2O + SO_2$ | $2.00 \times 10^{-16}$ |
| $O + S_3 \rightarrow SO + S_2$ | $8.00 \times 10^{-11}$ |
| $SO + S_3 \rightarrow S_2O + S_2$ | $1.00 \times 10^{-12}$ |
| $O + S_4 \rightarrow SO + S_3$ | $8.00 \times 10^{-11}$ |
| $O + SO \rightarrow SO_2$ | $5.30 \times 10^{-11}$ |
| $SO + O_3 \rightarrow O_2 + SO_2$ | $3.73 \times 10^{-17}$ |
| $SO + OH \rightarrow H + SO_2$ | $8.60 \times 10^{-11}$ |
| $SO + HO_2 \rightarrow OH + SO_2$ | $2.80 \times 10^{-11}$ |
| $SO + SO_3 \rightarrow SO_2 + SO_2$ | $2.00 \times 10^{-15}$ |
| $O + SO_2 + M \rightarrow SO_3 + M$ | $2.55 \times 10^{-35}$ |
| $OH + SO_2 \rightarrow HSO_3$ | $1.60 \times 10^{-12}$ |
| $HO_2 + SO_2 \rightarrow OH + SO_3$ | $1.00 \times 10^{-18}$ |
| $O + S_2O \rightarrow SO + SO$ | $1.54 \times 10^{-12}$ |
| $S_2O + S_2O \rightarrow S_3 + SO_2$ | $1.00 \times 10^{-14}$ |
| $SO_3 + H_2O \rightarrow H_2SO_4$ | $2.41 \times 10^{-15}$ |
| $SO_3 + OH \rightarrow HOSO_3$ | $3.92 \times 10^{-10}$ |
| $O_2 + HSO_3 \rightarrow HO_2 + SO_3$ | $4.79 \times 10^{-14}$ |
| $OH + H_2SO_4 \rightarrow H_2O + HOSO_3$ | $5.16 \times 10^{-14}$ |
| $HOSO_3 + O \rightarrow HO_2 + SO_3$ | $1.00 \times 10^{-12}$ |
| $HOSO_3 + OH \rightarrow H_2O + SO_2 + O_2$ | $1.00 \times 10^{-10}$ |

**Notes.**
[a] Units are $cm^3\ s^{-1}$ for two-body reactions and $cm^6\ s^{-1}$ for three-body reactions.
[b] Here, M is $H_2O$.

**Table 2**
The Upper Boundary Conditions of the Gas-phase Chemical Compositions in the Model

| Chemical Composition ($\times 10^{10}$ cm$^{-3}$) | H | O | $O_2$ | OH | $H_2$ | $O_3$ |
|---|---|---|---|---|---|---|
| Fixed density | 1.0 | 10 | 10000 | 0.2 | 10 | 500 |

**Table 3**
The Particle Energies of the Monoenergetic Beams and Their Estimated Penetration Depth

| Component | Average Energy (keV) | Estimated Penetration Depth ($\mu$m) |
|---|---|---|
| Thermal $H^+$ | 0.1 | 0.006 |
| Energetic $H^+$ | 600 | 10 |
| Thermal $O^+$ | 0.2 | 0.002 |
| Energetic $O^+$ | 300 | 1.3 |
| Thermal $S^{++}$ | 0.3 | 0.002 |
| Energetic $S^{++}$ | 300 | 0.8 |
| Electron | 500 | 1600 |

aforementioned model setup. These profiles are applied to the chemical-transport model as the driving forces of the chemical reactions in the surface ice. The number density profiles of the key nonwater species at each location are then simulated by our chemical-transport model. After that, we calculate the column densities of each species at each location to create the global distribution of the key species on the surface of Europa. The results are shown in Section 4.

### 4. Simulation Results and Sensitivity Tests

After analyzing the abundances of the nonwater species in our chemical-transport model, we notice that the surface ice of Europa is dominated by the following four species: $H_2SO_4$, $SO_2$, $H_2O_2$, and $O_2$. The abundances of other species are orders of magnitude lower than these four species. Therefore, we only discuss the global distribution of these four species in this section. Figures 2(a)–(d) illustrate the global distribution of $H_2SO_4$, $SO_2$, $H_2O_2$, and $O_2$, respectively. It should be noted that the column density of $H_2SO_4$ is about 2–3 orders of magnitude higher than the column densities of the other three species.

Figure 2(a) shows that $H_2SO_4$ has an eyeball-shaped distribution that is mostly concentrated in the low-latitude region of the trailing hemisphere of Europa. Such distribution is consistent with the Galileo observations of Europa on sulfuric acid and its hydrate (R. W. Carlson et al. 2005; J. B. Dalton et al. 2013). Compared with the energy flux distribution shown in Figure 1, we can see that the abundance of sulfuric acid is strongly correlated with the incoming particle flux of the thermal sulfur ions (see Figure 1(c)). Figure 2(b) shows that $SO_2$ is mostly concentrated in the low- to mid-latitude regions of the trailing hemisphere, but it is diluted in the high-latitude region and around the apex of the trailing hemisphere. The simulated $SO_2$ distribution is generally consistent with the mid-ultraviolet Hubble observations on $SO_2$ (see T. M. Becker et al. 2022). Compared with the global distribution of the incoming particle fluxes, we find that the $SO_2$ abundance is positively correlated with the incoming flux of the thermal sulfur ions and is further modulated by the electron energy flux. The extremely low electron energy flux around the apex of the trailing hemisphere accounts for the reduced $SO_2$ abundance in this region, whereas the small $SO_2$ peaks near the apex of the leading hemisphere correspond to the peaks in the distribution of the electron energy flux (see Figure 1(f)). As for the global distribution of oxidants, Figure 2(c) shows that $H_2O_2$ has a similar eyeball-shaped distribution as $H_2SO_4$, which is also concentrated in the trailing hemisphere. Such a distribution suggests that sulfur-related chemistry plays an important role in the production of $H_2O_2$. However, the simulated $H_2O_2$ distribution has a significant discrepancy with the observations (S. K. Trumbo et al. 2019; P. Wu et al. 2024), which indicate that $H_2O_2$ is concentrated in the leading hemisphere. The global distribution of $O_2$ on the surface of Europa is illustrated in Figure 2(d). The distribution of $O_2$ is strongly correlated to the distribution of the electron energy flux, which is mainly concentrated in the leading hemisphere and has a large cavity in the low- and mid-latitude regions of the trailing hemisphere. Currently there is no direct observation of $O_2$ column density in the ice of Europa to be compared with our simulation.

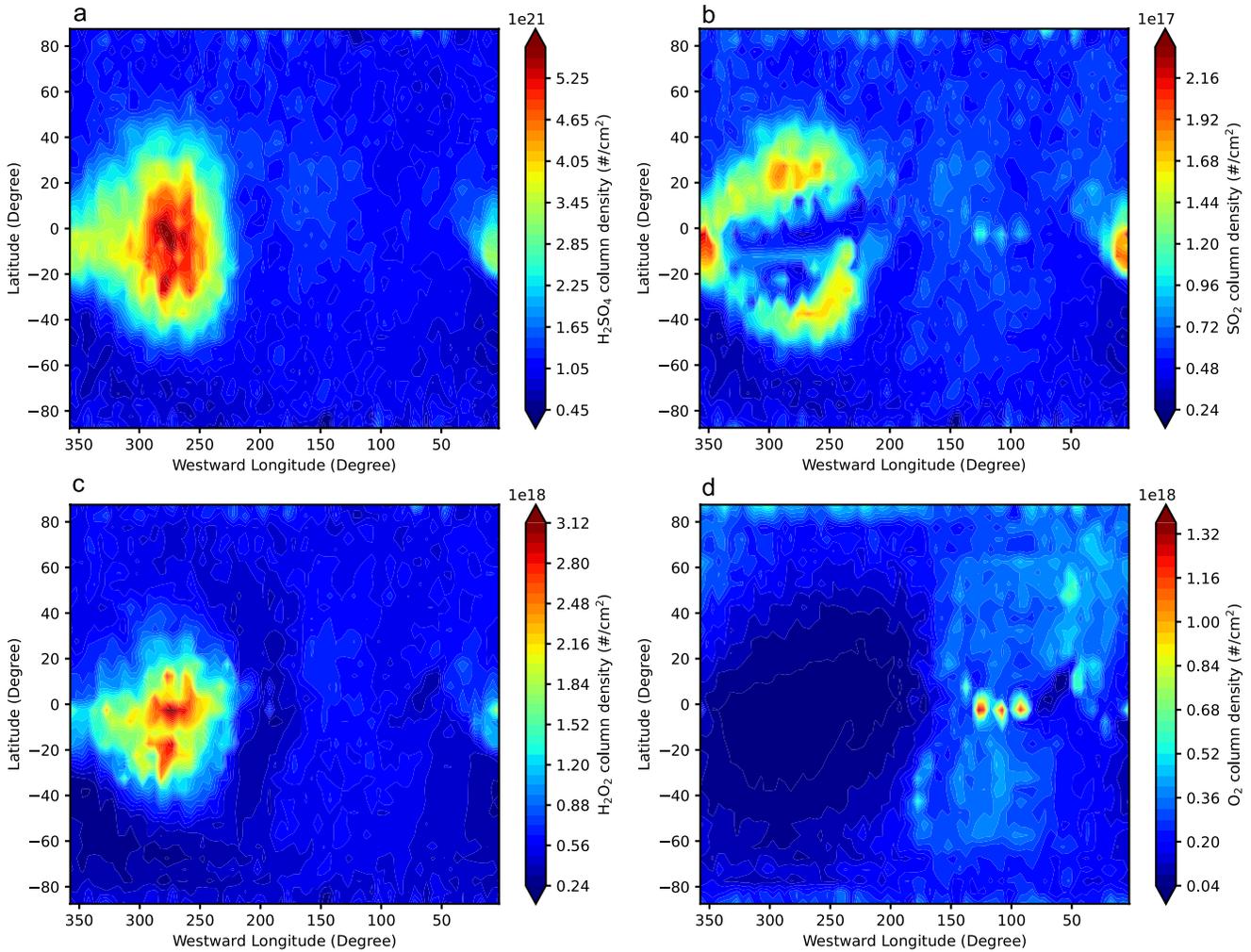

**Figure 2.** Global distribution of the four major nonwater species in the surface ice of Europa for the baseline model. (a) $H_2SO_4$. (b) $SO_2$. (c) $H_2O_2$. (d) $O_2$.

The diffusion coefficients and the reaction rate coefficients in the surface ice are two important sets of parameters in our model. However, the literature provides few constraints on these parameters. To evaluate the influence of these parameters on the simulation results, we conduct two test runs. In the first test run, we scale the diffusion coefficients of the gas-phase particles either down or up by a factor of 10. For both cases, the simulated distributions of the key species remain the same as in the baseline model, indicating that the simulation results are insensitive to the diffusion coefficients in the surface ice. In the second test run, we scale the reaction rate coefficients of all chemical reactions either down or up by a factor of 10. For the case that reaction rate coefficients are scaled up by a factor of 10, the simulated distribution of the four species remains consistent with those in Figure 2. However, when the coefficients are scaled down by a factor of 10, the global distributions changed significantly. Figure 3 shows the resulting distributions of the four key species. The eyeball-shaped distribution of $H_2SO_4$ (Figure 3(a)) remains the same as in Figure 2(a). As for $SO_2$, we find that the distribution of $SO_2$ becomes eyeball-shaped, and the reduced $SO_2$ abundance around the apex of the trailing hemisphere disappears. Such a distribution agrees well with the Hubble observations of $SO_2$ (T. M. Becker et al. 2022). The most significant change occurs in Figure 3(c), which illustrates the global distribution of $H_2O_2$. The maximum concentration of $H_2O_2$ appears in the low-latitude region of the leading hemisphere instead of the trailing hemisphere. The concentration of $H_2O_2$ becomes diluted in the trailing hemisphere, which agrees well with the $H_2O_2$ observations (S. K. Trumbo et al. 2019; P. Wu et al. 2024). The global distribution of oxygen molecules changes very little (see Figure 3(d)). These findings suggest that if we reduce the reaction rate coefficients in the ice, the modulation of electron flux on the abundance of $SO_2$ is weakened so that the low electron flux does not suppress the production of $SO_2$. They also suggest that the production of $H_2O_2$ becomes strongly correlated to the incoming electron flux and less correlated to the sulfur chemistry if the reaction rate coefficients are scaled down by a factor of 10. In our baseline model, we assume that the reaction rate coefficients in the ice are ~100 time lower than the rate coefficients measured in the gas phase (same as J. Li & C. Li 2023). Since Figure 3 agrees well with the observed global distribution of $H_2SO_4$, $SO_2$, and $H_2O_2$, we propose that lowering the gas-phase reaction rate coefficients by a factor of 1000 is a good estimation for the rate coefficients of the chemical reactions in the surface ice of Europa.

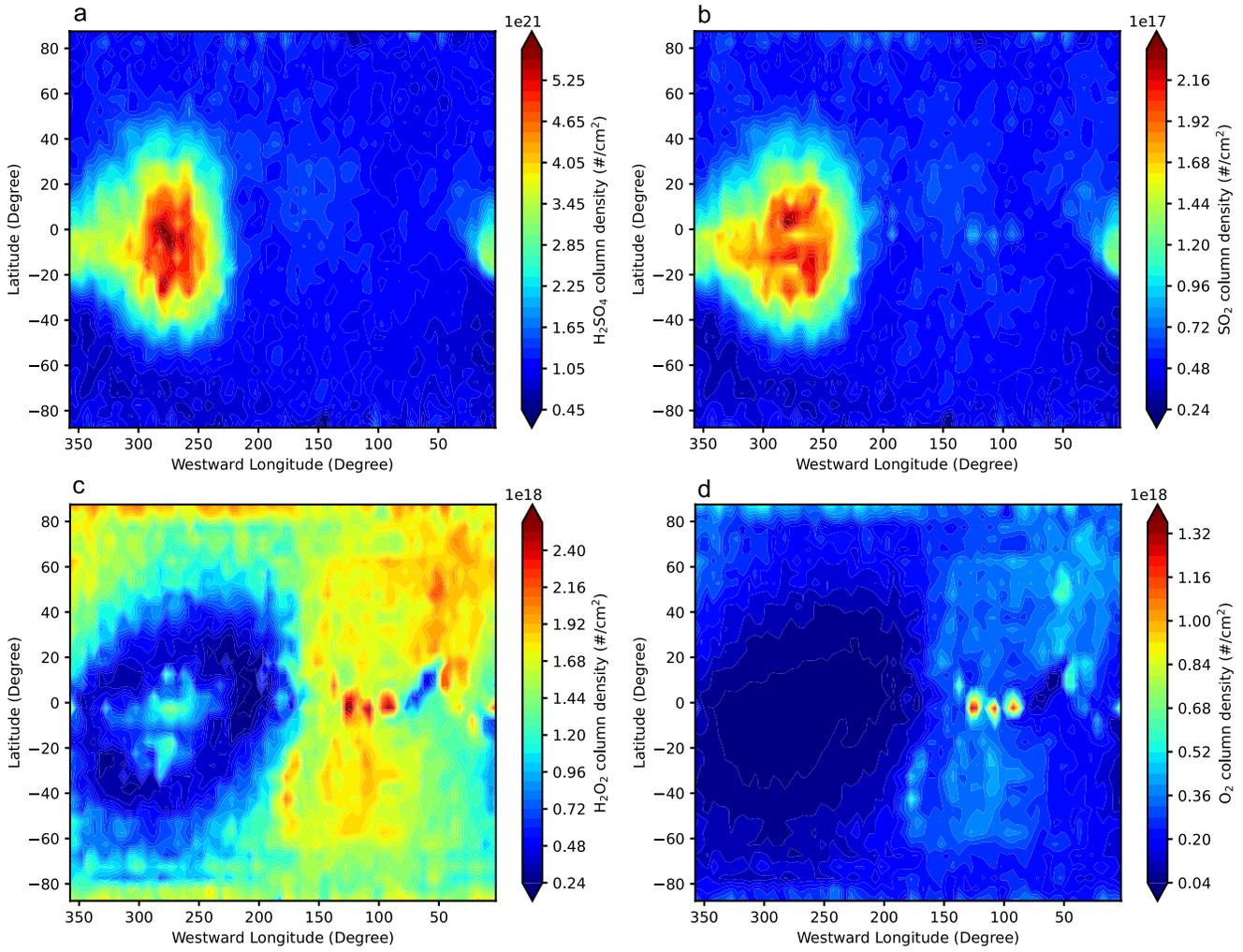

**Figure 3.** Global distribution of the four major nonwater species in Europa's surface ice with reaction rate coefficients scaled down by a factor of 10. (a) $H_2SO_4$. (b) $SO_2$. (c) $H_2O_2$. (d) $O_2$.

## 5. Conclusions and Discussion

Due to the impact of magnetospheric charged particles on the surface of Europa, the surface ice of Europa contains various nonwater chemical compositions. In this work, we make the first attempt to simulate the global distribution of the important nonwater species on the surface of Europa using the combination of a Jovian magnetospheric model and a chemical-transport model. The simulation results show the dominant nonwater compositions in the surface ice of Europa are $H_2SO_4$, $SO_2$, $H_2O_2$ and $O_2$, with $H_2SO_4$ being the most abundant composition. The initial results from our model partially reproduce the existing observations on the distributions of $H_2SO_4$, $SO_2$, and $H_2O_2$. When the reaction rate coefficients in the ice are reduced by a factor of 10, however, the model successfully reproduces all key features of the observations. This result provides a useful constraint on the reaction rate coefficients of chemical reactions in the ice. Furthermore, by comparing the simulated global distributions of the key species (Figure 3) with the incoming fluxes of major charged particle components from Jupiter's magnetosphere (Figure 1), we identify the following correlations: The abundances of $H_2SO_4$ and $SO_2$ are positively correlated to the particle flux of thermal sulfur ions. The abundance of $H_2O_2$ is strongly positively correlated to the electron flux and weakly positively correlated to the thermal sulfur ion flux. The abundance of $O_2$ is positively correlated to the electron flux. The simulated global distribution compensates for the limitations of the existing observations, which can only cover a portion of Europa, and addresses the lack of the key component, $O_2$. Although no direct observations of $O_2$ distribution on Europa's surface ice are available, W. M. Calvin et al. (1996) attributed absorption features at 573 and 627 nm on the icy moons to the trapped $O_2$ in the form of $O_2$–$O_2$ dimers. P. D. Cooper et al. (2003) proposed that $O_2$ production results from $H_2O_2$ decomposition, consistent with our finding that $O_2$ is primarily concentrated on the leading hemisphere, similar to $H_2O_2$. This study may also provide useful guidance to the future missions to Europa, e.g., Europa Clipper.

Although the simulation results demonstrate good consistency with the observations, there is one observed feature not fully captured in the simulation. The simulated abundance of $H_2O_2$ is considerably high in the polar regions, while the observations (S. K. Trumbo et al. 2019; P. Wu et al. 2024) suggest the abundance of $H_2O_2$ is relatively low in the polar regions. As a quantitative model, our model has the following uncertainties that may cause this discrepancy and other uncertainties in the simulated results:

1. To reduce the computational cost, the global distribution of the energy fluxes is obtained by averaging the energy fluxes on Europa of the three idealized simulation cases. Such a process may make the global distribution highly variable, especially for the electrons, which are quite variable and difficult to simulate in the plasma environment (shown in Figure 1(d)). This effect will make the simulated distribution of the incoming energy fluxes different from the actual distribution. Since some of the key species are sensitive to the energy fluxes of certain incoming charged particles, this process can bias our simulation results and cause the discrepancy.

2. In reality, each component of the magnetospheric charged particles has a certain energy spectrum. However, in our model, the incoming fluxes of the charged particles are simply treated as seven monoenergetic beams in the chemistry model. Meanwhile, the water dissociation and ion implantation processes are assumed to be uniformly happening along the penetration depth, but the actual profiles should be smoother than the step function used in our model (I. L. Barnett et al. 2012). In our model, the recombination rate due to the cage effect is assumed to be 50%, which may not accurately represent conditions on Europa. Variations in this rate would directly affect the water dissociation rate. These simplifications may make our simulated column densities different from the actual densities.

3. As mentioned in J. Li & C. Li (2023), most of the reaction rate coefficients used in this model are not directly measured at 100 K but are extrapolated based on the high-temperature rate coefficient measurements. Therefore, the actual rate coefficients at low temperatures may be different from the values used in our model. Also, fixing the temperature at 100 K for whole globe neglects the temperature difference between the polar region and the equatorial region. Since most reaction rates are temperature dependent and increase with rising temperature, the actual rate coefficients of most reactions are likely higher than those used in generating Figure 3. Consequently, the abundance of $H_2O_2$ in the polar region is probably lower than the value shown in Figure 3 and closer to that in Figure 2. Such an effect may account for the discrepancy between our simulated distribution and the observations. We plan to add a latitudinal temperature variation to our model in future studies.

4. S. M. Howell & R. T. Pappalardo (2018) suggest that the ice convection process may exist in the icy crust of Europa. This process can bring the materials from the subsurface ocean to the surface and effectively impact the surface processes of Europa and influence the abundances of the key nonwater species. Such an effect is not considered in the current model.

Considering that some of the limitations mentioned above are difficult to address and go beyond the scope of this article, we intend to explore these issues further in future research. Despite these limitations, this work provides the first simulation results on the global distribution of the key nonwater species on Europa. Comparison between the simulation results and the existing observations provides a constraint on the rate coefficient of the chemical reactions in the ice. This study also enhances our understanding of the global chemical processes of Europa, which is closely related to Europa's potential habitability. Since other icy satellites (e.g., Ganymede, Callisto, and Enceladus) are also embedded in high-radiation environments, our method can be applied to the surfaces of other icy satellites. The simulated global distribution of the key species may provide useful guidance to future missions to icy satellites, such as Europa Clipper and JUICE.

## Acknowledgments

J.L. is supported by the Science and Technology Development Fund (FDCT) of Macau (grant No. 002/2024/SKL). Support for Y.S. was provided by NASA grant No. 80NSSC20K0854 from the Solar System Workings Program. C.L. was supported by Heising-Simons Foundation and NASA's New Frontier Data Analysis Program grant No. 80NSSC23K0790.

## ORCID iDs

Jiazheng Li https://orcid.org/0000-0002-2563-6289
Yinsi Shou https://orcid.org/0000-0002-5765-9231
Cheng Li https://orcid.org/0000-0002-8280-3119
Xianzhe Jia https://orcid.org/0000-0002-8685-1484